\pdfoutput=1
\documentclass{JINST}
\usepackage{epsfig}
\usepackage{xspace}
\usepackage{longtable}
\usepackage{multirow}
\usepackage{graphicx}
\usepackage{booktabs}

\usepackage{fixltx2e}
\usepackage{amsmath,amssymb}
\usepackage[figuresright]{rotating}

\title{\bf 
A laser diode based system for calibration of fast time-of-flight detectors
}

\author{R. Bertoni, M. Bonesini~\footnote{Corresponding author.
E-mail:Maurizio.Bonesini@mib.infn.it} \\
Sezione INFN Milano Bicocca, Dipartimento di Fisica G. Occhialini,\\
Piazza
 Scienza 3, Milano, Italy \\
}
\author{A. de Bari, M. Rossella\\
Sezione INFN Pavia and Dipartimento di Fisica , via A. Bassi 6, Pavia, Italy}

\newpage
\abstract{
A system based on commercially available items, such as 
a  laser diode, emitting  in the visible range $\sim 400$ nm, 
and  multimode fiber 
patches, fused fiber splitters and optical switches may be assembled, 
for time calibration of multi-channels
time-of-flight (TOF) detectors with photomultipliers' (PMTs') readout.
As available  laser diode sources have unfortunately limited peak power,
the main experimental problem is 
the tight light power budget of such a  system.
In addition, while the  technology for fused fiber splitters
is common in the Telecom
wavelength range ($\lambda \sim 850, 1300-1500$ nm), it is not easily
available  in the visible one.
Therefore, extensive laboratory tests had to be  done on purpose,
to qualify the used optical components,  and a 
full scale   timing calibration prototype was built.
Obtained results show that with such a system, a calibration resolution
($\sigma$)  in the range 20-30 ps may be within reach.
Therefore, fast multi-channels TOF detectors, with timing resolutions
in the range 50-100 ps, may be easily calibrated in time.

Results on tested optical components may be of interest also for  time
calibration of different light detection systems based on PMTs, as 
the ones used for detection of the vacuum ultraviolet scintillation
light emitted by ionizing particles
in large LAr TPCs.  

}

\begin{document}

\section{Introduction and layout of the proposed calibration system }
\label{sec:Introduction}
Some examples of time-of-flight (TOF) detectors, based on scintillators 
with photomultipliers'(PMTs') readout and arranged along $x/y$ orthogonal coordinates,
are shown in table \ref{tof}. 
\begin{table}[hbtp]
\caption{
 Comparison of time resolutions for  TOF detectors, based on
 plastic scintillators with PMTs' readout.}
{\footnotesize
\begin{center}
\begin{tabular}{|c|c|c|c|c|}
\hline
Detector & Scintillator & $L \times W \times T (cm^3) $& PMT &
$\sigma_{t}$ (ps) \\
\hline MICE \cite{Bertoni:2010by} & BC404  or BC420 & 40-60 x 4-6 x 2.5 & Ham. R4998 & $\sim 50$ \\
\hline MEG \cite{Dussoni2010} & BC404 & 79.6 x 4 x 4 & Ham. 5924 & $\sim 60$ \\
\hline NA49 \cite{na49} & BC408 & 12(48) x 1 (1.25) x 1.5(2.4) & Ham. R3478 & 80 \\
\hline DASP \cite{dasp} & NE110 & 172 x 20  x 2 & RCA 8575 & 210 \\
\hline E813 \cite{e813} & BC408 & 200 x 8.5 x 5 & Ham. H1949 & 110 \\
\hline GlueX \cite{gluex} & Eljen  & 200 x 6.0 2.5 & XP2020 & $\sim 40$ \\
\hline PAMELA \cite{pamela} & BC404 & 41(18) x 33(15) x 07.(0.5) & Ham. 5900
 & 120 \\
\hline ARGUS \cite{argus}& NE110& 218 x 9.3 x 2 & RCA8575    & $\sim 210$ \\
\hline BELLE \cite{belle}& BC408 & 255 x 6 x 4 & Ham. 6680 & $\sim 100$ \\
\hline CDFII \cite{cdf} & BC408 & 279 x 4 x 4 & Ham. R7761 & $\sim 110$ \\
\hline CLEOII \cite{cleoII} & BC408 & 280 x 10 x 5 & XP2020 & 139 \\
\hline $N \overline{N}$ \cite{BaldoCeolin1992} & NE110 & 210-300 x 21 x 2 & XP2020 &
$\sim 300$ \\
\hline OBELIX \cite{obelix} & NE110 & 300 x 9.3x 4 & XP2020 & 170 \\
\hline E735 \cite{e735} &  BC408 & 305 x 10 x 5 & XP2020  & 110 \\
\hline MARKIII \cite{Brown1984} & NE Pilot F & 317.5 x 15.6 x 5.1 & XP2020 & $\sim 171$ \\
\hline CLAS \cite{Smith1999} & BC408 & 32-450 x 15-22 x 5.1 & XP4312B/D1 & 163 \\
\hline DIRAC \cite{dirac} & BC420 & 40 x 7 x 2.2 & Ham. R1828-01 & 123 \\
\hline TOPAZ \cite{Kishida1987} & Eljen  & 400 x 13 x 2.5 & Ham. H1949 & $\sim 200$ \\
\hline HARP\cite{BaldoCeolin2004}             & BC408 & 180-250 x 21 x 2.5 & XP2020 & $\sim 160$ \\
\hline
\end{tabular}
\label{tof}
\end{center}}
\end{table}

For a particle crossing
an individual  scintillation counter $i$ ,
with double-side PMTs readout ($j=1,2$),
the time difference $\delta t_{i,j}$ between the STOP signal from
the PMT $j$  and the START signal from a reference counter $t_s$ is given by:
\begin{equation}
\delta t_{i,j}=t_0 + \frac{L/2 \pm x}{v_{eff}}-t_s + \delta_{i,j}
 \ \ \ \ \ \ \ \ 
 j=1,2 
\label{eq:deltat}
\end{equation}
where $t_0$ is the particle crossing time, $x$ its distance from
the counter center, {\em L} is the scintillator length, $v_{eff}$ the 
effective velocity of light 
in the scintillator  ($v_{eff}^{-1}\sim 6.2$ ns/m) and
 $\delta_{i,j}$ includes
all system delays (cables, PMT transit time, etc.).

To assure optimal performances, it is essential to 
determine precisely the individual channel delays $\delta_{i,j}$. They may drift during
data-taking, due to temperatures excursions and other effects. 
As reported in reference \cite{BaldoCeolin2004}, 
standard RG58 signal cables  have 
time variation up to 
95 ppm/${}^\circ$C, due to thermal excursions, while  the better RG213 
cables reach values around 30 ppm/${}^\circ$C. 
With typical time delays around 100 ns, time drift of the order of 9 (3) 
ps/${}^\circ$C may be reached by using  RG58 (RG213) signal cables. This has to
be compared with a TOF detector resolution ($\sigma_t$) in the range 50-150 ps.

The quantity
\begin{equation}
\widehat{\delta t_{+,i}}=\frac{\widehat{\Delta t_{i,1}} + 
\widehat{\Delta t_{i,2}}}{2}=t_0+\frac{L}{2 
\cdot v_{eff}}-t_s
\label{media}
\end{equation}
where $\widehat{\delta t_{i,j}}$ are the time differences $\delta t_{i,j}$ 
corrected for the system delays $\delta_{i,j}$,
does not depend on the particle impact point and
allows the measurement of its time-of-flight  
(TOF). This may be used for particle identification (PID).


Time calibration of a TOF  system means the precise
determination of delays $\delta_{i,j}(t)$ at a start time $t_0$ and the monitoring
of their  change along the data-taking period. 
Cosmic rays, as in the case of the HARP Tof Wall \cite{BaldoCeolin2004},
 or impinging beam particles, as in the case of the  
MICE TOF system \cite{Bertoni:2010by}, may be used.
Another way is to deliver  fast calibration pulses 
to each individual channel. The requirement on the light calibration pulses is 
that their rising edge mantains the original time characteristics with
minimal deterioration,
up to the injection point in the scintillation counter  
and that their time delays do not sensibly vary during the calibration 
procedure.

For a TOF system, made of two detectors, of which the first gives the 
start signal ($t_s$), the TOF measurement resolution is given by:
\begin{equation}
\sigma_{TOF}=\sqrt{\sigma_{T_1}^2+\sigma_{T_2}^2+\sigma_{cal}^2}
\label{eq3}
\end{equation}
where $\sigma_{T_i} (i=1,2)$ is the intrinsic resolution of each TOF detector 
and $\sigma_{cal}$ the resolution of the calibration system, used to determine the delays.

The request on the calibration system is to have $\sigma_{cal}$ as small as 
possible. This translates, in practice, into the request to have calibration 
pulses with the smallest width ($\sigma_{laser \ pulse ^{t}}$) 
and thus the best rising edge determination. The intrinsic laser width must be 
minimal and the additional 
spread introduced by the laser pulse delivery system kept as
small as possible. If possible, all must be realized with components 
commercially
available for easiness.

Laser based calibration systems were used
for the $N-\overline{N}$ experiment
at Grenoble \cite{BaldoCeolin1992}, the MARK-III experiment at SLAC 
\cite{Brown1984},
the CLAS system at CEBAF \cite{Smith1999}, the TOPAZ experiment at KEK
\cite{Kishida1987} and the HARP/PS214 experiment at CERN \cite{Bonesini2003}.

As an example, 
in reference \cite{Bonesini2003} a custom-made 
duplicated Nd/Yag laser  at 532 nm, 
with passive Q-switch
and active/passive mode locking was used~\footnote{model SYLP0 from 
Quanta Systems srl, Italy, with 60 ps FWHM, 10 Hz repetition rate, 
3 mJ energy per pulse}. It was followed by a custom-made
 optical delivery system to 
the individual
scintillators, based on a bundle of 64 Corning SMF-28 IR monomode 
fibers~\footnote{made by Fiberlan srl, Milano. The fibers in the bundle 
behave as ``limited number of modes'' fibers at the wavelength of interest
($\sim 532$ nm), with a measured dispersion of $\sim 3.6 ps/m$.}. 
Unfortunately, a laser of this type is
expensive and  difficult to operate. 

By using this system, a calibration resolution ($\sigma_{calib}$) around 70 ps 
was quoted for the large Tof Wall of the HARP experiment at CERN PS 
\cite{BaldoCeolin2004}.  

All the previous systems comply with the previous requests
 only in part, especially
for the use of commercially availabile optical components and laser systems.

The use of turnkey low-cost laser-diode systems, but unluckily with limited
peak power, has been proposed  by the T0 detector group of the ALICE 
Collaboration at LHC \cite{Bondila2005}. 
Laser diodes have high repetition
rates (up to several MHz) but regrettably 
low energies per pulse (up to a factor 
$10^{-6}$ lower, as respect to conventional systems), putting severe constraints on the optical laser pulse
delivery system to the individual channels in terms of attenuation.

A TOF detector laser diode calibration system may be built up from optical 
switches
that direct the input laser pulse to one of $N$ output channels, fused
fiber splitters $ 1 \times M$  that divide the input laser pulse to $M$ 
output channels and fiber patch cables for connections between the 
previous items~\footnote{Such a calibration system may be used also for 
timing systems 
based on PMTs different from conventional scintillation time-of-flight
detectors, such as PMTs systems used for the detection of the vacuum ultraviolet(VUV) light emitted by ionizing particles in large LAr TPCs. In this case
requirements on the time calibration system may be  relaxed, as
required time resolutions are in the {\it ns}  range, instead that in the 
50-150 {\it ps} one.}

\begin{figure}[hbt]
  \centering
  \includegraphics*[width=\linewidth]{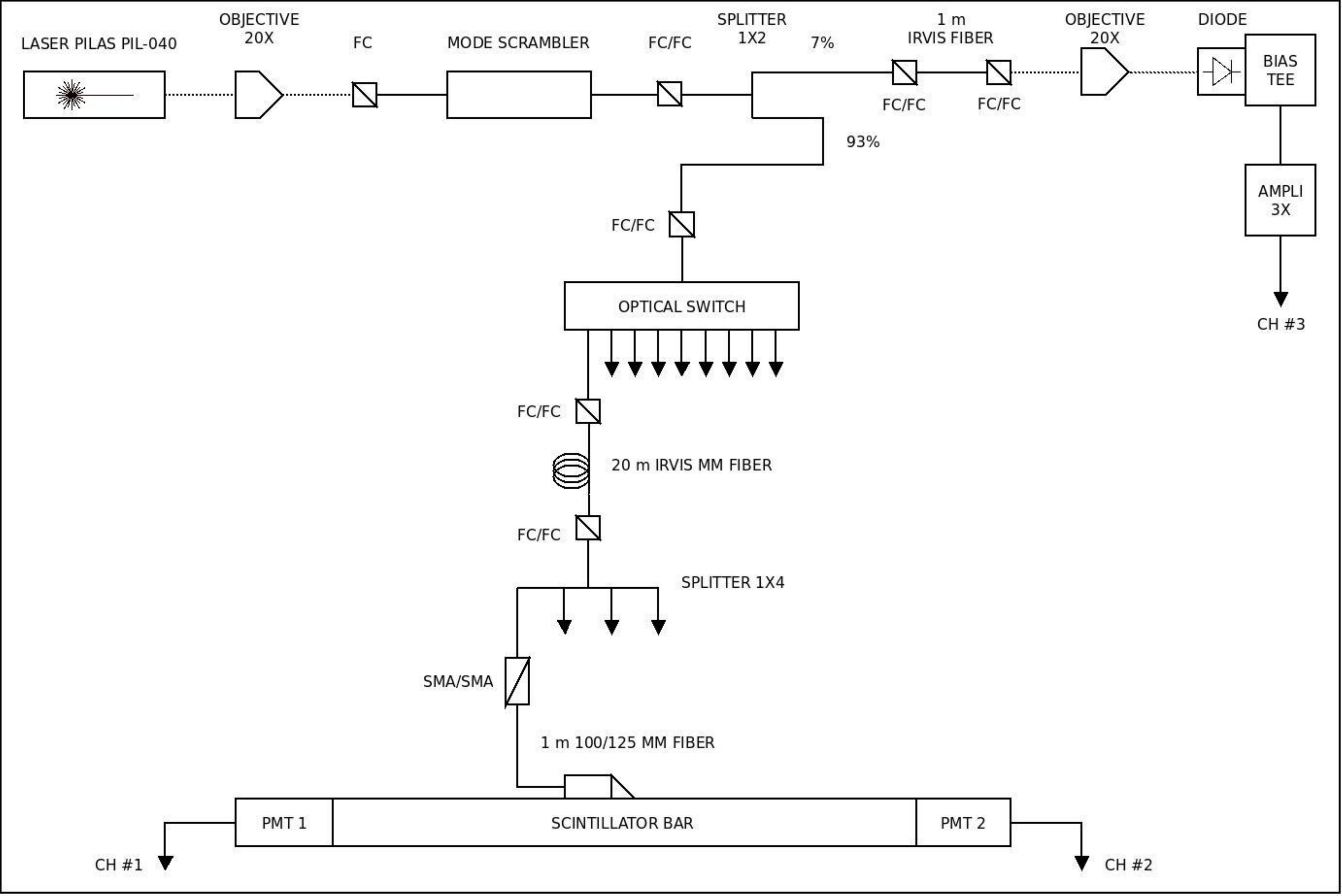}
  \caption{Schematic layout of the tested prototype calibration system.}
  \label{fig:setup}
\end{figure}

A simple layout is shown in figure \ref{fig:setup} 
that illustrates the prototype
system developed at INFN Milano Bicocca. 
The start signal ($t_s$) is given by a fast photodiode (ch 3), while the stop
signal is given by the signals from the two PMTs at the end of the counter
scintillator bar under test (ch no 1,2).

The aim of this paper was to develop a calibration system for fast TOF
detectors based on commercially available optical components, avoiding 
custom-made solutions, and using low-cost fast diodes. The main concern were
to keep the timing characteristics of the delivered calibration 
laser pulses under 
control (by reducing $\sigma_{laser \ pulse}^{t}$ to a minimum) 
and to manage efficiently 
the system power budget, to allow calibration of a multi-channel system
(up to some hundreds). 
 
The setup used to characterize the optical components is described in
section \ref{sec:tests}, while the optical characterization of elements
is shown in section \ref{sec:comp} and the estimated 
prototype performances are shown
in section \ref{sec:proto}.

\section{Experimental test setup}
\label{sec:tests}
\begin{figure}
  \centering
 \includegraphics*[width=0.8\linewidth]{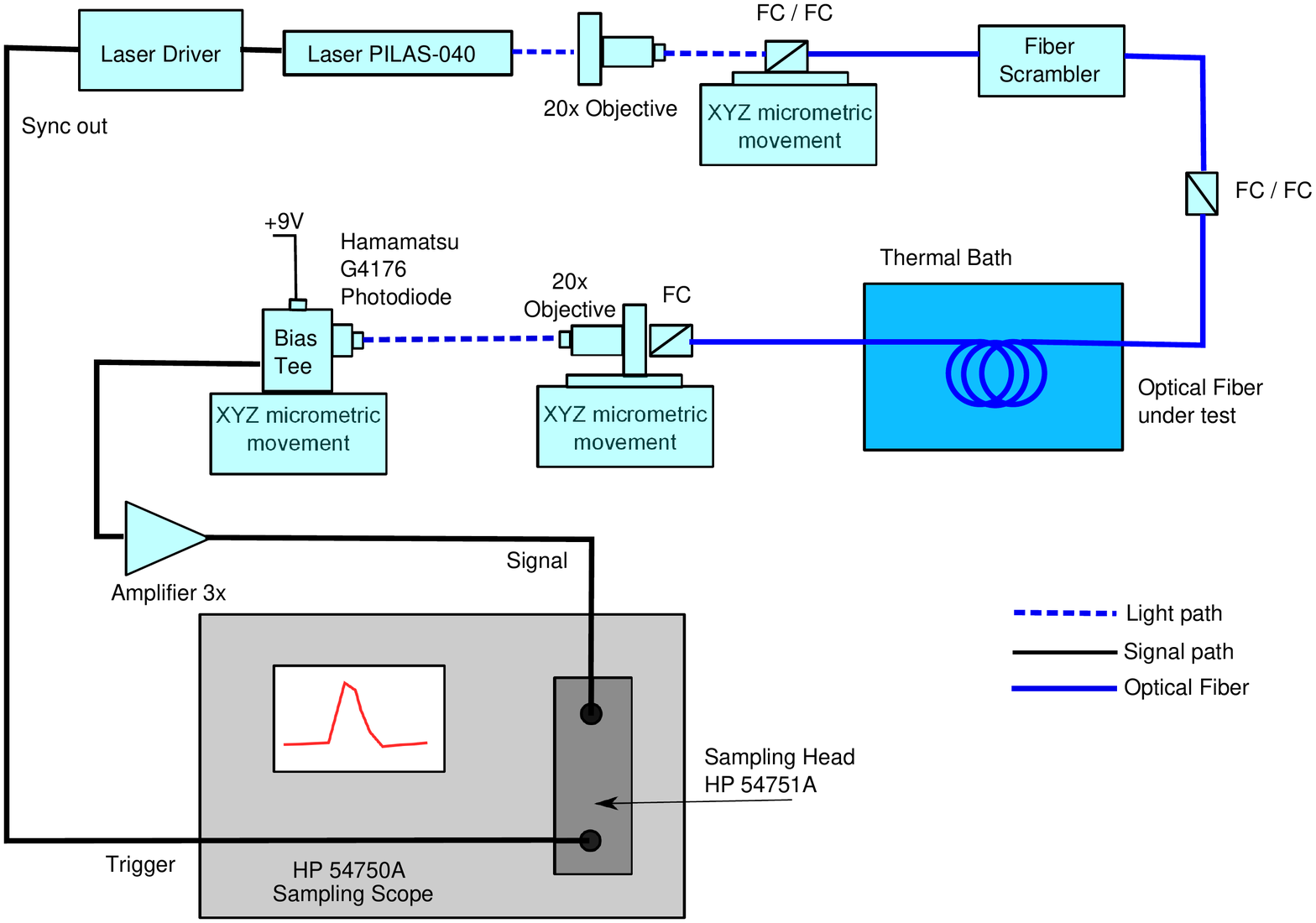}
  \caption{Layout of the optical components test setup. The 
    Hamamatsu G4176 photodetector+HP54750 sampling scope was replaced 
by an OPHIR 
Nova  powermeter, in the attenuation measurements. In some measurements, 
the optical components under test were put inside 
the thermal bath provided by a   Lauda cooling thermostat RP485.}
  \label{fig:laser}
\end{figure}

The layout of the experimental setup used for the measurement of the 
optical components under test is shown in figure 
\ref{fig:laser}. 
Light pulses from a fast laser~\footnote{ Model Pilas 040 from Advanced 
Laser Systems, with FWHM $\sim 30$ ps, repetition rate up to 1 MHz, peak
power 1 W, free space beam head optics} are injected into a
 Multimode (MM) Arden Photonics ModCon mode scrambler, with $\leq 3$ dB 
insertion loss at 850 nm \footnote{
A mode scrambler is used to reduce dependency from the light source in
bandwith and attenuation measurements for multimode (MM) fibers, giving a more 
uniform injection into the fiber}  (or into a 1m MM fiber) using an Olympus 
20x microscope objective, with 0.4 N/A and 1.2 mm working distance. Fine alignment for 
the injection is obtained via
a $x,y,z$ Newport manual micrometric stage. 
After the last optical element under 
test, the output light pulse is focalized, via another 20x microscope 
objective placed on
a micrometric 3-axis Thorlabs flexure stage, on a fast 
InGaAs MSM Hamamatsu G4176 photodiode (with 30 ps rise and fall time). 
For optimal performances, the photodetector is powered by a 1 GHz broadband 
Picosecond Pulse Lab 5550B bias tee~\footnote{ thus providing the bias
voltage to the active device, while allowing high speed signals to pass
through with minimum signal degradation}   and its
signal is  amplified by a 10dB broadband inverting amplifier~\footnote{
Model BBA-3 from Alphalas Gmbh, with 15 GHz bandwidth}. The amplified
signal is then measured with a 20 GHz HP 54750 sampling scope, where the 
trigger signal is given by the laser sync out. 

The laser source  has been 
characterized by the manufacturer with a 20 GHz Ultrafast detector,
with FWHM < 20 ps, and a  50 GHz HP54750A scope.
The timing characteristics of the injection system used are shown 
in table \ref{tab:las}. Our results compare well with 
datasheet FWHM data for G4176 photodetectors from
the manufacturer, that are in the range 70-80 ps with a similar test
setup.

\begin{table}[hbt]
\centering
\caption{Measured timing characteristics of the used laser injection
system. An Arden Photoncs mode scrambler (or a 1m OZ/Optics IRVIS fiber)
was used in the measurements. Individual FWHM's are added in quadrature.}
\begin{tabular}{|c|c|c|}
\hline
                & expected $FWHM_0$  & measured $FWHM_0$ \\ \hline
laser           & $ \sim 35$ ps           &                    \\
bias tee        & $ \sim 30$ ps           &                    \\
HP54750 scope   & $\sim 22$ ps           &                    \\
G4176 photodiode &  $\leq 55$ ps  ($\simeq 42$ ps typ.)  &    \\
fast amplifier   & $\sim 30$ ps          &                    \\ \hline
total            & $\leq 81$ ps ($\simeq 73$ ps typ.) & $80.5 \pm 0.3$ ps (with MS) \\
                 &                                    & $78.7 \pm 0.1$ ps (without MS)\\
\hline    
\end{tabular}
\label{tab:las}
\end{table}

Timing characterization of single optical components, such as fiber patch 
cords, optical switches, fused fiber splitter, has been done by putting
them after the laser light injection stage and measuring the increase
in the signal FWHM ($\Delta FWHM= \sqrt{ FWHM^2 - FWHM^2_0}$).  
From the measured FWHM, one can then compute $\Delta \sigma_t$ using a gaussian
approximation for the signal shape. In most cases, this is true aside for 
the longer fiber patches ($\simeq \ 20$ m) where non gaussian tails appear in
the falling edge of the signal. 

As their timing characteristics are determined by measuring the increase of 
the FWHM, as respect to $FWHM_0$, the precise determination of $FWHM_0$ 
is essential. From the reported mesurements a relative error around a few 
per mille may be quoted. 

\section{Tests of optical system components}
\label{sec:comp}

A critical point for a time calibration 
system is the stability in time of the used laser.
Tests were done with the available Pilas 040 laser, at 10 KHz repetition
rate, using a 10 m long MM fiber as a delay before the photodetector.
On a timescale of several hours of continuos operation, the measured $10-90 \%$
risetime had a maximum variation of $ 0.4 $ ps (6 per mille effect) 
and the pulse
time delay  ($\sim 120 $ ns) of less than 12 ps ($< 0.1 $ per mille effect).
 
For an optimal behaviour of the calibration system, in all optical elements 
the laser pulse must be transmitted with minimimal 
attenuation and mantaining its timing characteristics, 
mainly its risetime.
Assuming a gaussian shape for the impulse, 
the key parameter for
characterisation of such waveform is its standard deviation $\sigma_t$
~\footnote{We remind here for easiness the main relations between 
often measured time parameters in the gaussian approximation: 
$FWHM = 2.354 \times \sigma_t$, $ risetime (10-90 \%)= 1.6888 \times \sigma_{t} =
falltime (10-90\%) $}.

For each optical component under test a complete set of measurements was 
done to  characterize both its transmission properties (power measured with the
optical powermeter or  pulse amplitude, $V_{max}$, measured 
via the sampling scope) and its timing properties ($10-90 \%$ risetime, 
$10-90 \%$ falltime, FWHM and pulse delay).
Preliminary results were reported in reference \cite{tipp2014}.  

\subsection{Characterization of used optical fibers}

To guarantee an optimal and simple injection of the light from the laser
source, large core multimode fibers (MM) are to be preferred to small core
single mode (SM) fibers. The problem is that MM fibers may suffer from
 a remarkable
deterioration of the timing properties of the propagating laser pulse, due to
modal dispersion. 
This has to be checked in the real experimental conditions 
on the distances
of interest  for the calibration setup, e.g. with fiber
patchcords up to 15-20 meters, to go from the laser source to the 
detector channels to be calibrated. 
Tests were done with the setup of figure \ref{fig:laser} where the signal
at the end of  the fiber under test was measured directly by 
a powermeter~\footnote{Model Ophir Nova with a PD300 head} for attenuation 
studies and by the full setup with a Hamamatsu G4176 photodetector and a 
HP54750 sampling scope for the timing properties studies.
Attenuation studies were done with and without  the optical mode scrambler.
\begin{figure}[hbt]
  \centering
  \includegraphics*[width=0.49\linewidth]{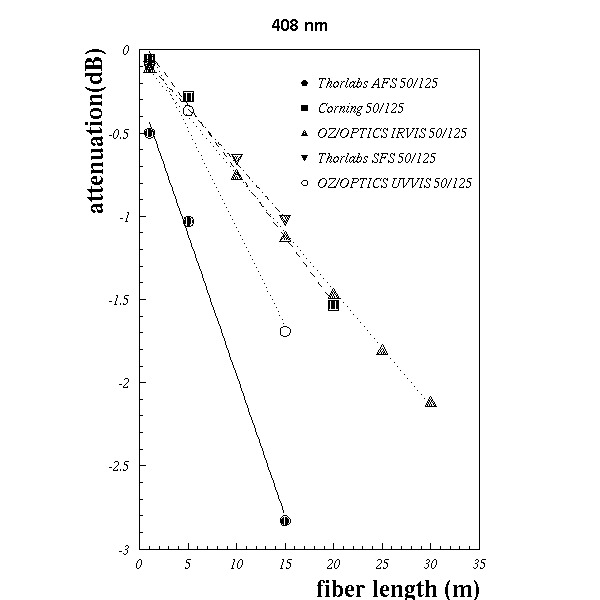}
  \includegraphics*[width=0.49\linewidth]{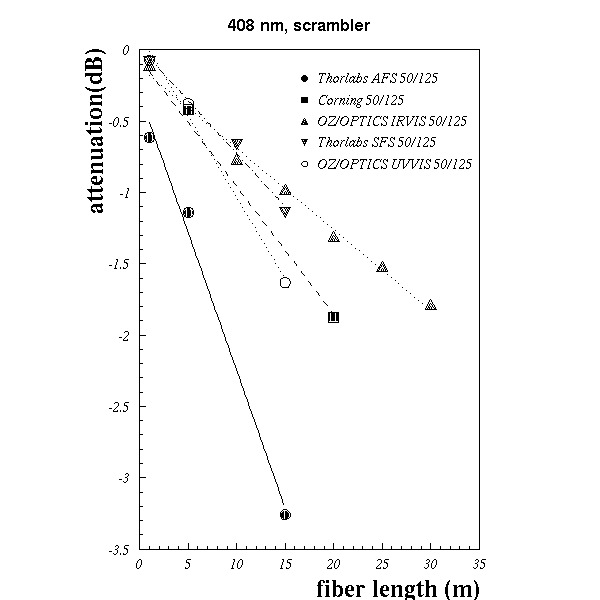}
  \caption{Attenuation in dB for the different types of 50 $\mu m$ core
optical fibers under test. 
   Left panel: tests without mode scrambler. Right panel: tests with mode scrambler.}
  \label{fig:att}
\end{figure}

\begin{figure}[hbt]
  \centering
  \includegraphics*[width=0.49\linewidth]{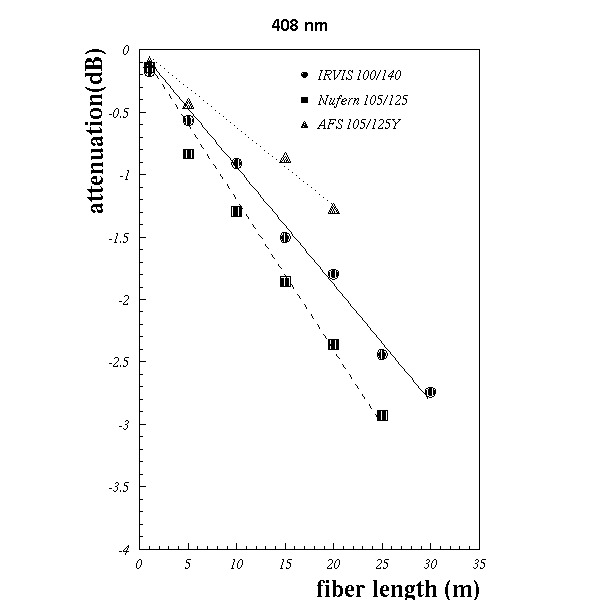}
  \includegraphics*[width=0.49\linewidth]{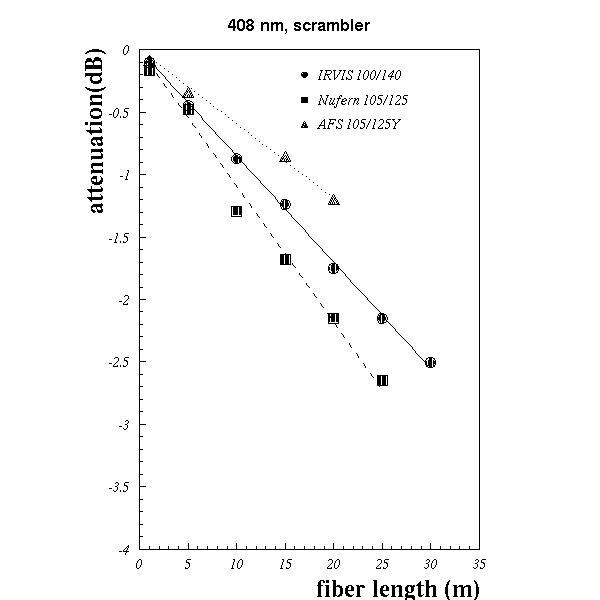}
  \caption{Attenuation in dB for the different types of 100 $\mu m$ core
optical fibers under test. 
   Left panel: tests without mode scrambler. Right panel: tests with mode scrambler.}
  \label{fig:att1}
\end{figure}

Figure \ref{fig:att} and table \ref{tab:res} show
 the attenuation measured in decibel (dB) with different 
types of $50 \mu m$ MM core fibers, using patches of different lengths.
Results for $100 \mu m$ core fibers are  shown  in figure 
\ref{fig:att1} and  reported also in table \ref{tab:res}.
Measurements are affected by systematic errors mainly due to the
fluctuations in the laser intensity (within $1 \%$). 

\begin{figure}[hbt]
  \centering
  \includegraphics*[width=0.7\linewidth]{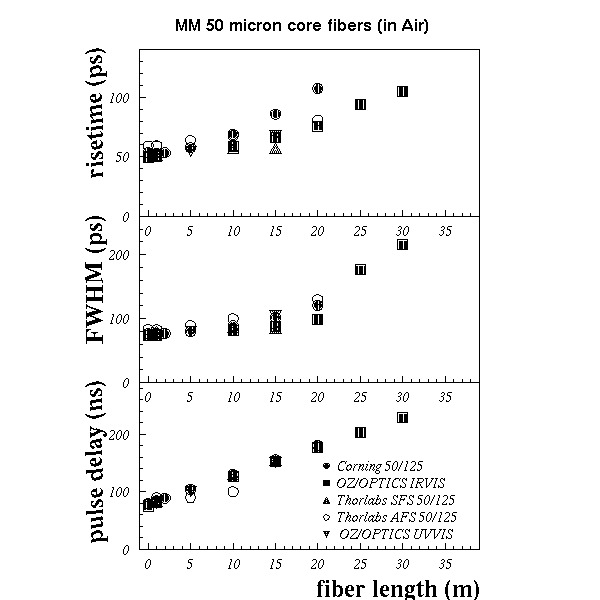}
  \caption{Increase in 10-90 \% risetime (top panel), FWHM
   (middle panel) and pulse delay (bottom panel) inserting a fiber patch cord 
   of length L after the input insertion mode scrambler. Results are for all the tested 50 $\mu$m core MM fibers.}
  \label{fig:disp}
\end{figure}

\begin{figure}[hbt]
  \centering
  \includegraphics*[width=0.7\linewidth]{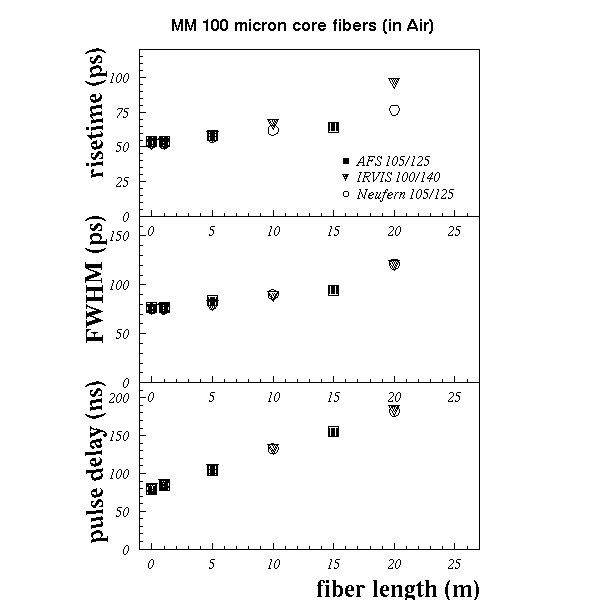}
  \caption{Increase in 10-90 \% risetime (top panel), FWHM
   (middle panel) and pulse delay (bottom panel) inserting a fiber patch cord 
   of length L after the input insertion mode scrambler. Results are for 
   different $100 \mu$m core fibers.}
  \label{fig:disp1}
\end{figure}

\begin{table}[hbt]
\centering
\caption{Attenuation properties of measured 50 and 100 $ \mu$m core MM fibers.
Measurements are affected mainly by the systematics arising from laser 
intensity fluctuations.}
\vspace{.2cm}
\begin{tabular}{|c|c|c|c|}
\hline
MM fiber  &  fiber type & Attenuation(dB/m) & 
Attenuation (dB/m)  \\
          &             & (no scrambler) & (scrambler) \\ \hline
Thorlabs AFS 50/125Y & step index: 400-2400 nm & $0.169 \pm 
0.007$ & $0.193 \pm 0.011 $ \\
                     & 50 $\mu m$ core; 0.22 NA & & \\
Thorlabs SFS 50/125Y & step index: 250-1200 nm & $0.066 \pm 
0.005$ & $0.074 \pm 0.010 $ \\
                     & 50 $\mu m$ core; 0.22 NA & & \\
Corning 50/125 & graded index & $0.079 \pm 
0.003 $ & $0.089 \pm 0.005 $ \\
                     & 50 $\mu m$ core; 0.20 NA & & \\
OZ/OPTICS  50/125 IRVIS& graded index: 400-1800 nm & $0.070 \pm 
0.003$ & $0.057 \pm 0.001 $ \\
                     & 50 $\mu m$ core; 0.20 NA & & \\
OZ/OPTICS  50/125 UVVIS& step index: 200-900 nm & $0.118 \pm 
0.009$ & $0.114 \pm 0.007 $ \\
                     & 50 $\mu m$ core; 0.22 NA & & \\ \hline
 Nufern S105/125-22A        & step index: 800 - 1600 nm &  $0.120 \pm 0.003$ & 
$0.109 \pm 0.003$ \\
                    & 100 $\mu m$ core; 0.22 NA & & \\ 
 Thorlabs AFS 105/125Y & step index: 400-2400 nm & $0.063 \pm 
 0.039$ & $0.060 \pm 0.039 $ \\
                     & 100 $\mu m$ core; 0.22 NA & & \\ 
 OZ/OPTICS  100/140 IRVIS& graded index: 400 - 1800  nm & $0.094 \pm 
 0.002$ & $0.085 \pm 0.002  $ \\
                     & 100 $\mu m$ core; 0.29 NA & & \\ 
\hline
\end{tabular}
\label{tab:res}
\end{table}
Results on pulse dispersion, in terms of pulse delay and increase 
in the 10-90 \% risetime or in the FWHM  are shown in figure
\ref{fig:disp} for  several $50 \mu$mm core MM fiber patches. All tests were done 
with a mode scrambler before the fiber patch to be tested. 
Additional results for 100 $\mu$ m core diameter  fiber patches
are shown instead in figure \ref{fig:disp1}. 

Most fiber patches under test have FC (``Fiber Channel'') connectors, 
that behave better than
SMA (``Subminiature A'' microwave) connectors traditionally used for MM fibers. 
The influence of the 
insertion of commercially available FC/FC mate sleeves was studied with
specimen from different producers. The only information directly provided by 
the producer is their insertion loss (attenuation). 
Results with a Corning 
50/125Y 10 m long fiber were compared with the ones with two 5 m fibers, 
of the same type, using in between a FC/FC mate sleeve. No appreciable 
difference was seen, as shown in table \ref{tab:FC}, aside a small change
in the pulse delay ($\sim 1$ per mille). 

\begin{table}[hbt]
\centering
\caption{Effects of FC/FC mate sleeves. Specimen A is from OZ/Optics, B and
C from Lightel}
\begin{tabular}{|c|c|c|c|}
\hline
patch  & $V_{max}$ (mV)  & FWHM (ps) & pulse delay (ns)   \\ \hline
10 m           & 76.73 $\pm$ 0.25 & 89.72 $\pm$ 1.95  & 129.710 $\pm$ 0.002 \\
5m + 5m mate A & 76.46 $\pm$ 0.23 & 91.10 $\pm$ 1.80  & 129.580 $\pm$ 0.002 \\
5m + 5m mate B & 76.07 $\pm$ 0.19 & 89.42 $\pm$ 0.90  & 129.590 $\pm$ 0.002 \\
5m + 5m mate C & 75.95 $\pm$ 0.18 & 89.58 $\pm$ 1.29  & 129.590 $\pm$ 0.003 \\ 
\hline
\end{tabular}
\label{tab:FC}
\end{table}

\subsection{Characterization of used optical switches and fused fiber splitters}

Optical switches direct the input light signal to one of $N$ outputs 
with minimal insertion losses. The maximum number of available
output channels is typically 16-32 in the visibile range, using 
MM fibers. Optical fiber fused splitters instead split the input 
signal over $N$ output channels in an even way (if required). These
last components are quite common for the Telecom range of wavelengths
(850 nm or 1300-1500 nm) but difficult to find for the visible range
at $\sim 400$ nm. 

After the mode scrambler, 
a MM optical switch $1 \times 9$ made by PiezoJena GmBH~\footnote{
Model F-109-05 with Thorlabs SFS 50/125 fibers as pigtails} was put.
The measured FWHM of the laser signal 
increased from $80.47 \pm 0.61 $ ps to  
$83.14 \pm 0.46$ ps. 
At 408 nm, the output signal variation from channel to channel was within
1 \%, with a cross-talk $\sim 2\%$ and an insertion loss
$\sim 2.2$ dB. These numbers are to be compared with 
manufacturer specs, at $\lambda = 850 $ nm, where
an insertion loss of 1.5 dB and a cross-talk $\sim 70$ dB 
were  quoted. 

\begin{figure}[hbt]
  \centering
  \includegraphics*[width=0.49\linewidth]{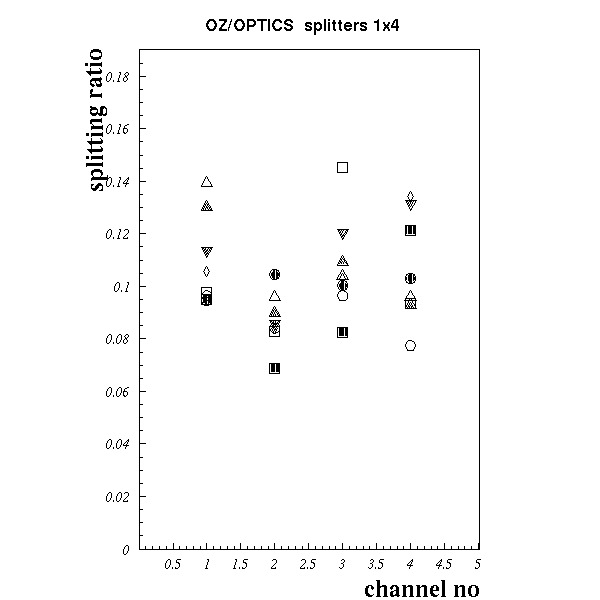}
  \includegraphics*[width=0.49\linewidth]{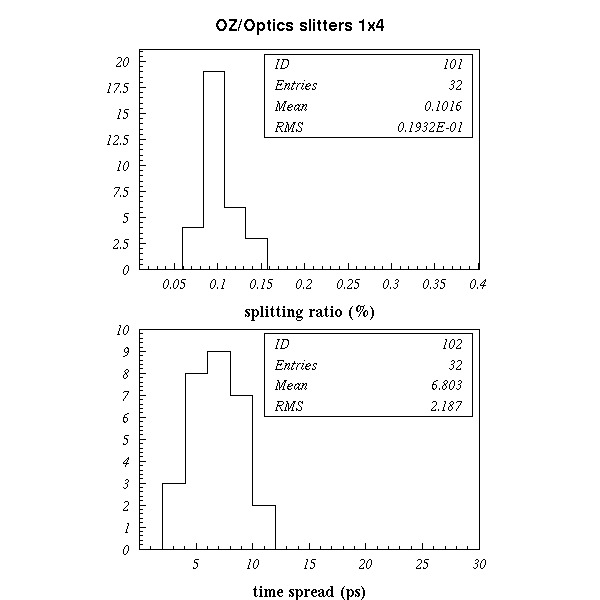}
  \caption{Left panel: splitting ratio vs channel number for some 
   1 $\times$ 4 OZ/Optics splitters, made of 50 $\mu$m core fibers. 
Each different symbol used (such as $\triangle, \Box, \bullet$, ...) refers to 
one of the
eight specimen of the OZ/Optics splitters under test. 
Right panel: for all the output channels of theOZ/Optics
 $1 \times 4$ splitters under test, distribution of splitting ratio (\%) 
and introduced time spread ($\Delta \sigma_{t}$).}
  \label{fig:split1}
\end{figure}

\begin{figure}[hbt]
  \centering
  \includegraphics*[width=0.49\linewidth]{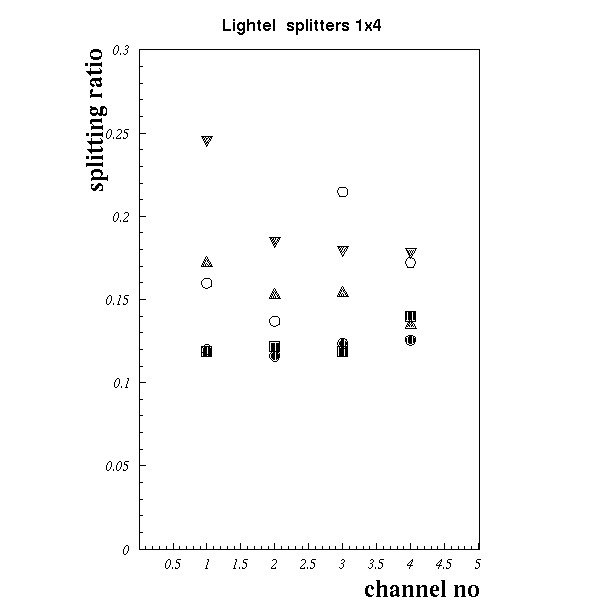}
  \includegraphics*[width=0.49\linewidth]{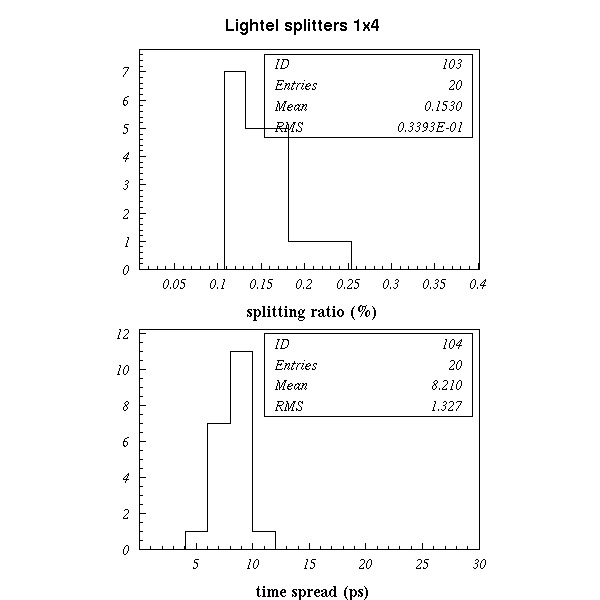}
  \caption{Left panel: splitting ratio vs channel number for some 1 $\times$ 4
   Lightel
   splitters, made of 50 $\mu$m core fibers. 
Each different symbol used refers to 
one of the four specimen of the Lightel splitters under test. 
Right panel: for all the output channels of the Lightel $1 \times 4$ 
splitters under test, distribution of splitting ratio (\%) and introduced
time spread  ($\Delta \sigma_{t}$).}
  \label{fig:split2}
\end{figure}

\begin{figure}[hbt
]
  \centering
  \includegraphics*[width=0.49\linewidth]{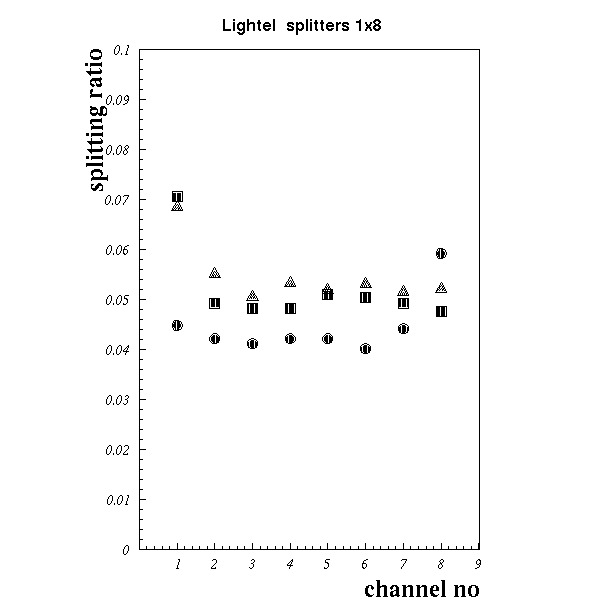}
  \includegraphics*[width=0.49\linewidth]{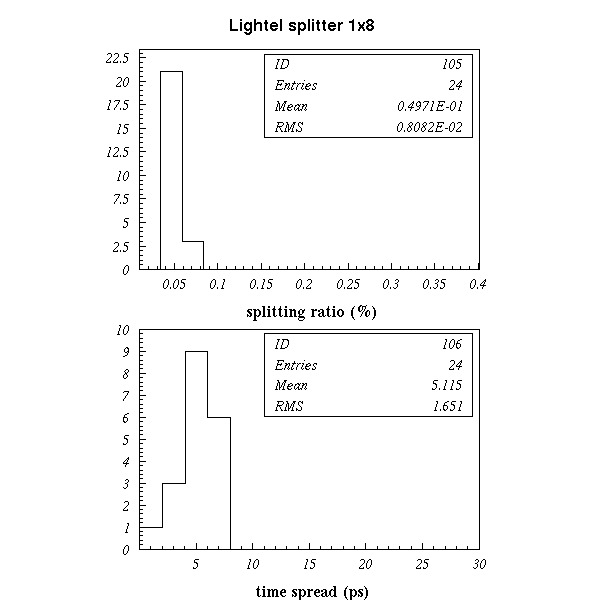}
  \caption{Left panel: splitting ratio vs channel number for a 1 $\times$ 8
Lightel splitter, made of 50 $\mu$m core fibers. 
Each different symbol used  refers to 
one of the three specimen of the Lightel  splitters under test. 
Right panel: for all the output channels of the Lightel $1 \times 8$ 
splitters under test, distribution of splitting ratio (\%) and introduced
time spread ($\Delta \sigma_{t}$).}
  \label{fig:split3}
\end{figure}

Several  optical splitters $1 \times 2$, $1 \times 4$, 
$1 \times 8$   made  by OZ/Optics and Lightel,
with 50 $\mu$m or 100 $\mu$m core fibers, were tested. 
The  optical fused fiber splitters replaced
the optical switch, after the mode scrambler in the test setup. A relevant 
insertion loss at 400 nm  (around 2-3 dB) and a dispersion in the splitting ratio 
$\sim 10-15 \%$ was seen. The measured FWHM of the input laser signal
increased typically of $2-3$ ps  after insertion of the optical splitters.  
 Figures \ref{fig:split1} to \ref{fig:split3}  show the output splitting ratio 
(in \%)~\footnote{defined as the power ratio between the total input and a
single channel output}  and the increase in the pulse timing resolution 
for some $1 \times 4$ and $1 \times 8$ splitters from OZ/Optics and Lightel, 
made with 50$\mu$m fibers. Results for splitters made both with 
50 $\mu$m or 100 $\mu$ 
core fibers are resumed instead in table \ref{tab:split1}.
Results are reported as average values $\pm$ rms values to give
an idea of the distribution width over the specimens of the considered 
samples~\footnote{ Spreads in some cases are large as specimen were from
different manufacturer's production batches. Uniformity of response 
may be a major concern for proper operations of passive fused splitters}. 
Entries for the third and fourth columns of table \ref{tab:split1}
were obtained from distributions, as the ones reported in figures 
\ref{fig:split1} to \ref{fig:split3}.



\begin{table}[hbt]  
\centering
\caption{Measured properties of MM fused fiber splitters from OZ/Optics and
Lightel Ltd.
Results are reported as average values $\pm$ rms values.
The total insertion loss (in dB) is defined at the power ratio
between the sum of the output channels and the input one.} 
\vspace{.2cm}
\begin{tabular}{|c|c|c|c|c|}
\hline
  & no of tested  & splitting ratio   & $\Delta(\sigma_t) (ps)$ & tot. insertion loss (dB)  \\ 
  & specimens               &                   &                    & \\ \hline
OZ/Optics 1x4 50 $\mu$m  & 8 & $0.10 \pm 0.02$ & $6.80 \pm 2.19$ & $4.49 \pm 0.71$  \\
Lightel 1x4 50 $\mu$m    & 5 & $0.15 \pm 0.03$ & $8.21 \pm 1.33$ & $2.29 \pm 0.77 $\\
Lightel 1x8 50 $\mu$m    & 3 & $0.05 \pm 0.01$ & $5.12 \pm 1.65$  & $3.95 \pm 0.48 $\\
OZ/Optics 1x4 100 $\mu$m & 2 & $0.21 \pm 0.03$ & $6.65 \pm 1.55 $ & $0.80 \pm 0.02$ \\
Lightel 1x4 100 $\mu$m   & 2 & $0.22 \pm 0.02$ & $4.51 \pm 2.78$ & $0.63 \pm 0.09$ \\
\hline
\end{tabular}
\label{tab:split1}
\end{table}

\subsection{Temperature dependence for optical components}

\begin{figure}[hbt]
  \centering
  \includegraphics*[width=0.60\linewidth]{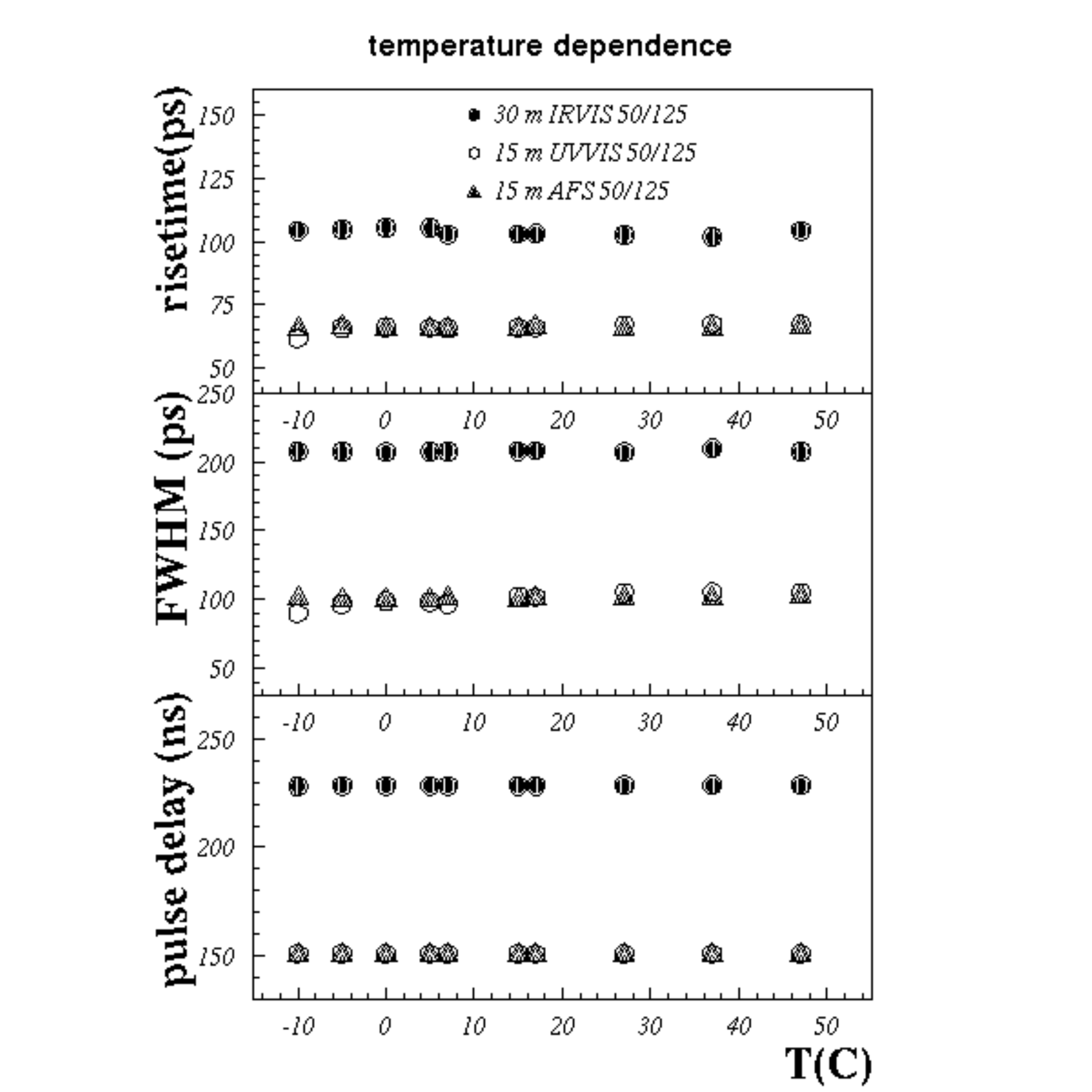}
  \caption{Temperature dependence for different MM fiber patches.
  10-90\% risetime, FWHM, pulse delay and signal pulse height (P.H.) 
  vs temperature. }
  \label{fig:temp}
\end{figure}

As the variation of delays $\delta_{i,j}$ is mainly due to thermal excursions
in the experimental hall housing the used TOF system, it is 
important to study also the influence of temperature on the  
fibers to be used in the calibration system. 
A precision LAUDA cooling thermostat RP845
 (precision $\pm 0.01  \ ^0 C$ of the thermal bath), where a part of the 
fiber under test was  kept at fixed 
temperature, was added to our test system. Different patches were tested: 
two 15 m Thorlabs AFS
50/125 and OZ/OPTICS UVVIS fiber 
patches (14 m in the thermal bath, 1 m in air) and one 30 m 
OZ/OPTICS IRVIS 50/125 fiber patch (28 m in the thermal bath, 2m in air). Results are shown
in figure \ref{fig:temp} and no relevant influence is seen in the operating range
between -10 and 50 $^{0}C$ for the most important  parameters.  

\begin{figure}[hbt]
  \centering
  \includegraphics*[width=0.59\linewidth]{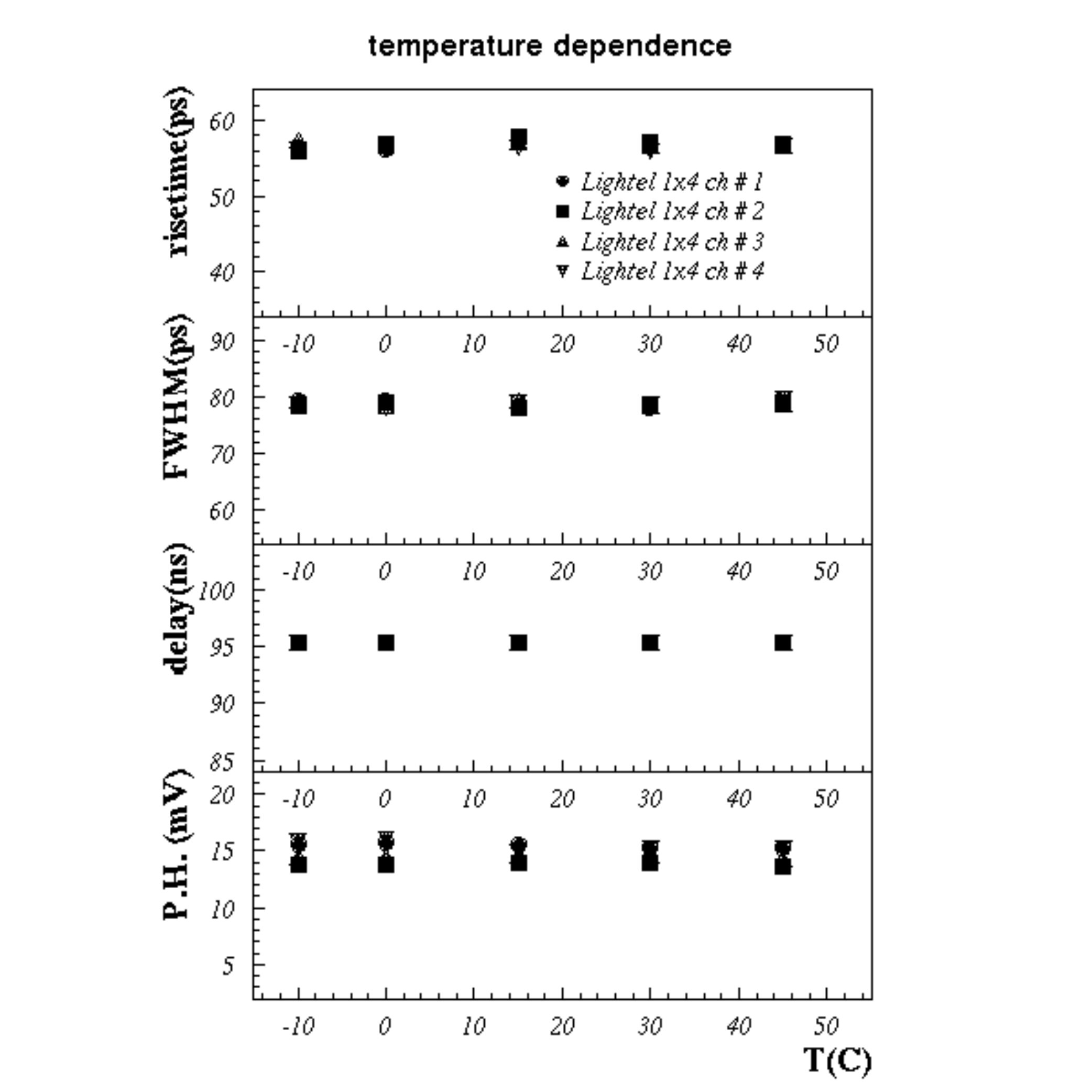}
  \includegraphics*[width=0.59\linewidth]{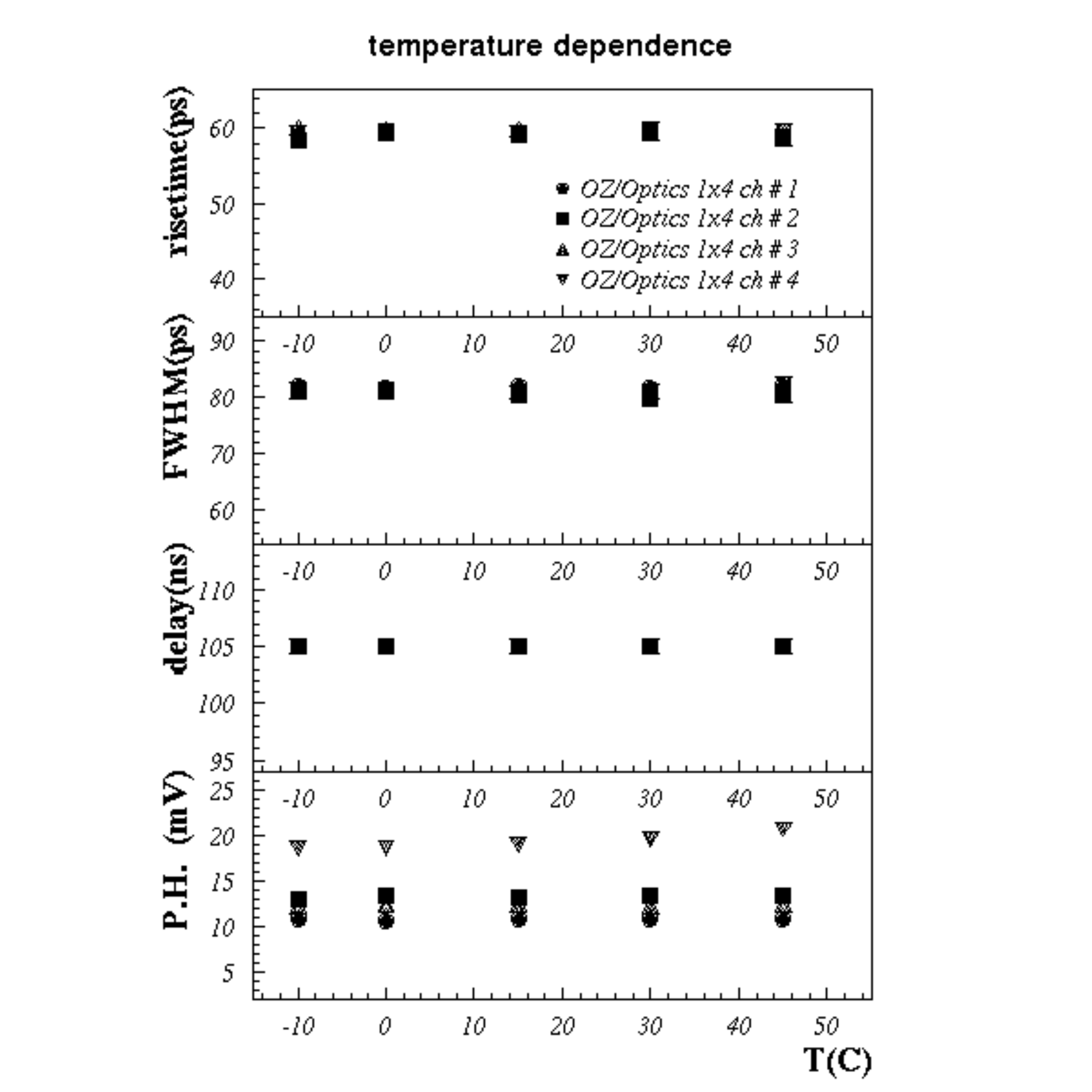}
  \caption{Temperature dependence for two typical  1x4 50 micron fibers 
           fused splitters: one Lightel 1x4 (top panel), one OZ/Optics 1x4 
           (bottom panel). Results are reported for the four different output 
            channels of a splitter. }
  \label{fig:temp1}
\end{figure}

The same tests were done to measure the behaviour of a specimen of 
Lightel 1x4 and
OZ/Optics 1x4 optical splitters as a function of temperature. Results are shown in figure \ref{fig:temp1}  and, as
before,  no relevant influence 
is seen in the operating range between -10 and 50 $^{0}C$.

In conclusion, the tested optical components: switches, fused fiber splitters, fiber
patchcords seem all suitable to assemble the pulse delivery system for 
a diode laser 
based calibration system, either with 50 $\mu$m or 100 $\mu$m MM fiber's 
core diameter, mantaining the original time characteristics of the input
laser calibration pulse.

\subsection{Timing properties of the full laser pulse delivery system}  

A laser pulse delivery system, 
based on optical switches, fused optical splitters 
and fiber patches to inject light into the channels to be
calibrated, provides light calibration pulses with time resolution
given by:  
\begin{equation}
\sigma_{laser \ pulse}^{t}=\sqrt{\sigma_{laser}^2+\sigma_{switch}^2+\sigma_{splitter}^2
+\sigma_{fiber \ patches}^2}
\label{eq2}
\end{equation}
where $\sigma_{laser}(\sim 15-30)$ ps is the original laser pulse resolution, 
$\sigma_{switch} (\sim 5-10)$ ps  is the time
spread introduced by the optical switch, $\sigma_{splitters} (\sim 5-10)$ ps
is the one
introduced by the fiber splitters and $\sigma_{fiber \ patches} (\sim 1-2$ ps/m) 
is the one  
introduced by the used fiber patches.

In the prototype calibration system, assembled  at INFN
Sezione Milano Bicocca, OZ/OPTICS IRVIS 50 $\mu$m fibers were used. They have
a pulse width dispersion $\sim 1 $ ps/m, an attenuation $\sim 0.06$ dB/m, 
with a temperature effect $ < 0.2 \%$ assuming
an temperature excursion of $10 \  ^{0}C$ for a 10 m fiber, and a delay 
$\sim 5.11$ ns/m.  
In its present configuration, up to 36 (72) individual channels (using $1 \times 4$
or $1 \times 8$ splitters) may be calibrated. It may be easily extended to configurations
with up to 100-200 channels. Configurations with more channels require more powerful
diode-lasers systems, that are difficult to find now on the market or a more tight
control on the power budget of the system. 
Figure \ref{fig:setup} shows the setup presently assembled in laboratory, where the
light from one $1 \times 4$ splitter is injected, by means of a reflection prism, 
into the center of the scintillation counter to be calibrated.  

Using the test setup of figure \ref{fig:laser}, where the full laser pulse 
delivery system 
under test (up to the splitter output) is put between the fiber scrambler and the Hamamatsu G4176 
photodiode, it was possible to estimate the timing characteristics of the 
proposed system. 
They are resumed in table \ref{tabx}. 
From table \ref{tabx} we may estimate that, for the system under test, 
$\sigma_{laser \ pulse} \sim 35$ ps in agreement with the estimate from 
equation \ref{eq2}~\footnote{in this estimate $\sigma_{laser \ pulse}$ is
evaluated through the quadratic difference of FWHM at the end of the 
calibration system and after the Mode Scrambler, taking into account 
our measurement of the intrinsic laser width}.  

\begin{table}[hbt]
\centering
\caption{Main characteristics of the laser pulse delivery
 system under test. Results
at end of the fiber splitters are averaged over the output channels.
P.H. is the signal pulse height in mV, as measured by the digital
sampling scope.}
\vspace{.2cm}
\begin{tabular}{|c|c|c|c|}
\hline
          & after Mode Scrambler & at end of cal. system &
at end of cal. system  \\ 
          &                      & (OZ/Optics 1x4 splitters)  &
(Lightel 1x4 splitters)  \\ \hline
P.H. (mV) & $69.4 \pm 0.2$ & $1.9 \pm 0.7$ & $1.8 \pm 0.2 $ \\
risetime (ps)   & $56.3 \pm 0.5$ & $ 105.3 \pm 6.0$ & $98.2 \pm 1.7 $ \\
FWHM (ps) & $80.5 \pm 0.3 $ & $118.3 \pm 9.0 $ & $122.3 \pm 1.5 $ \\ 
\hline
\end{tabular}
\label{tabx}
\end{table}

 
\section{Test of a prototype calibration system}
\label{sec:proto}

The calibration system  resolution ($\sigma_{cal}$) is determined by the  
goodness of the 
timing properties of the rising edge of the laser calibration pulse~\footnote{
strongly dependent of the smallness of the laser pulse width, that in the
gaussian approximation is related to the rising edge via the formula:
risetime(10-90 \%) = 0.717 x FWHM}, that gives the 
STOP signal, and from the jitter on the START signal (in our case given
by the fast photodiode in ch no 3). Therefore, only a direct test may estimate
properly the value of the calibration resolution.
    
The prototype calibration system, shown in figure 
\ref{fig:setup}, was tested using  a 1m long 100/125 $\mu$m MM fiber to inject light into the
scintillator bar of the system to be calibrated.

About $4-6\%$ of the laser light arrives to the injection prism for the TOF
scintillator bar under test, depending on the type of $1 \times 4$ optical splitter 
used. 
With a trigger on cosmic rays, put at the center of the scintillation counter, it was
possible to see that the calibration signal in  the counter is roughly a factor 2-3
bigger than the one of  cosmic muons. 

The PMTs' signals were acquired with a VME system, based on a CAEN 
V2718 interface,
and  sent after a 50\% passive splitter to a CAEN V792 charge integrating 
analog-to-digital module (QADC) and a CAEN V1290 time-to-digital module (TDC), 
after a CAEN V895 leading edge discriminator, with a threshold set 
at a -50 mV value. 
The two TDC signals ($TDC_L, TDC_R$) are computed as differences between 
the L/R PMT discriminated 
signal (STOP) and a reference signal (START) sent, after the $1 \times 2$
 splitter, 
to a fast 
photodiode~\footnote{Thorlabs DET02A, rise time 50 ps and fall time 150 ps,
with CAEN A1423 wideband amplifier}. 

A home-written data acquisition system acquired data from both TDC's and ADC's
as a binary file, that was later analyzed with the ROOT package \cite{root}.

\begin{figure}[hbt]
  \centering
  \includegraphics*[width=0.45\linewidth]{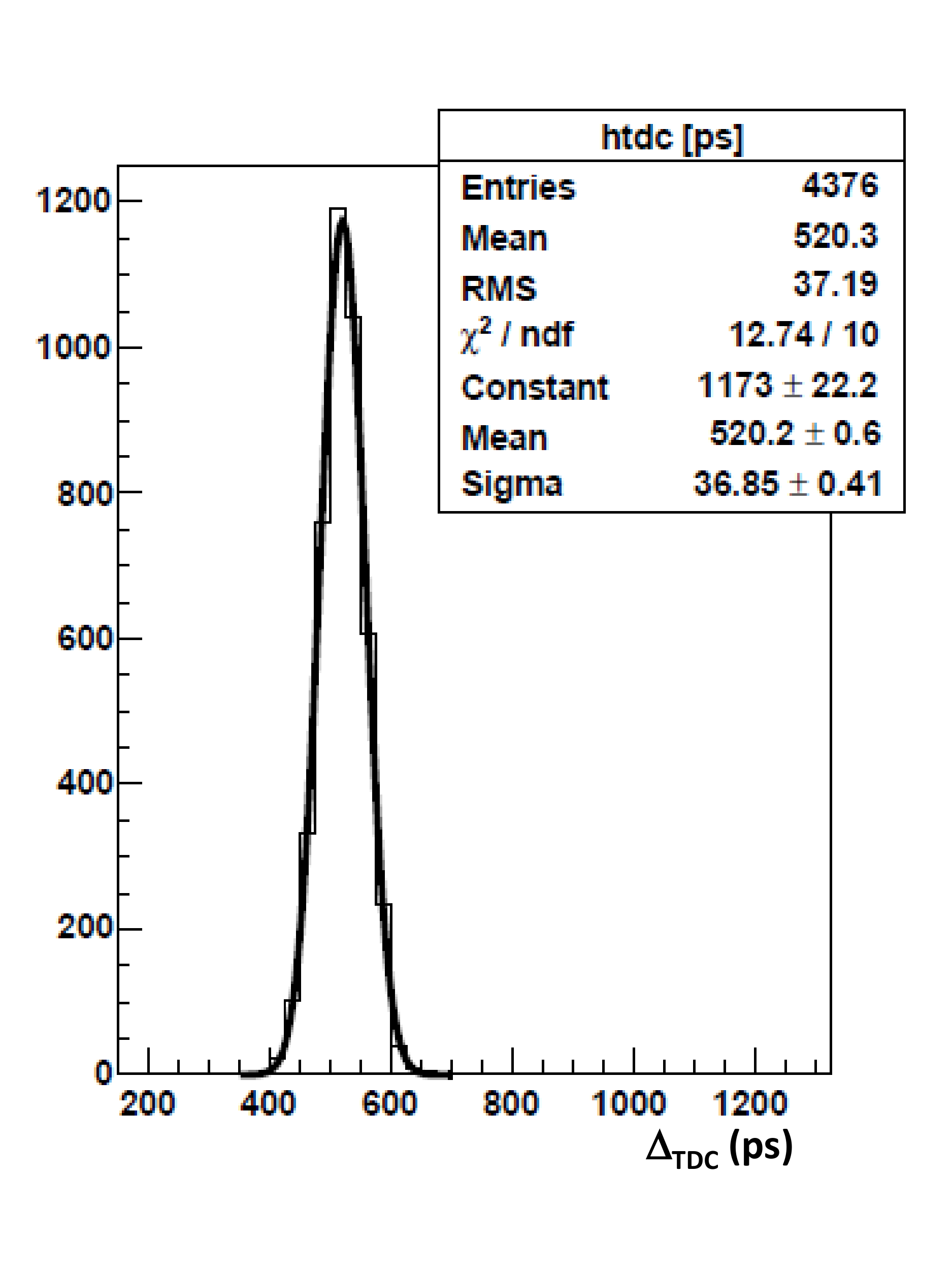}
  \includegraphics*[width=0.50\linewidth]{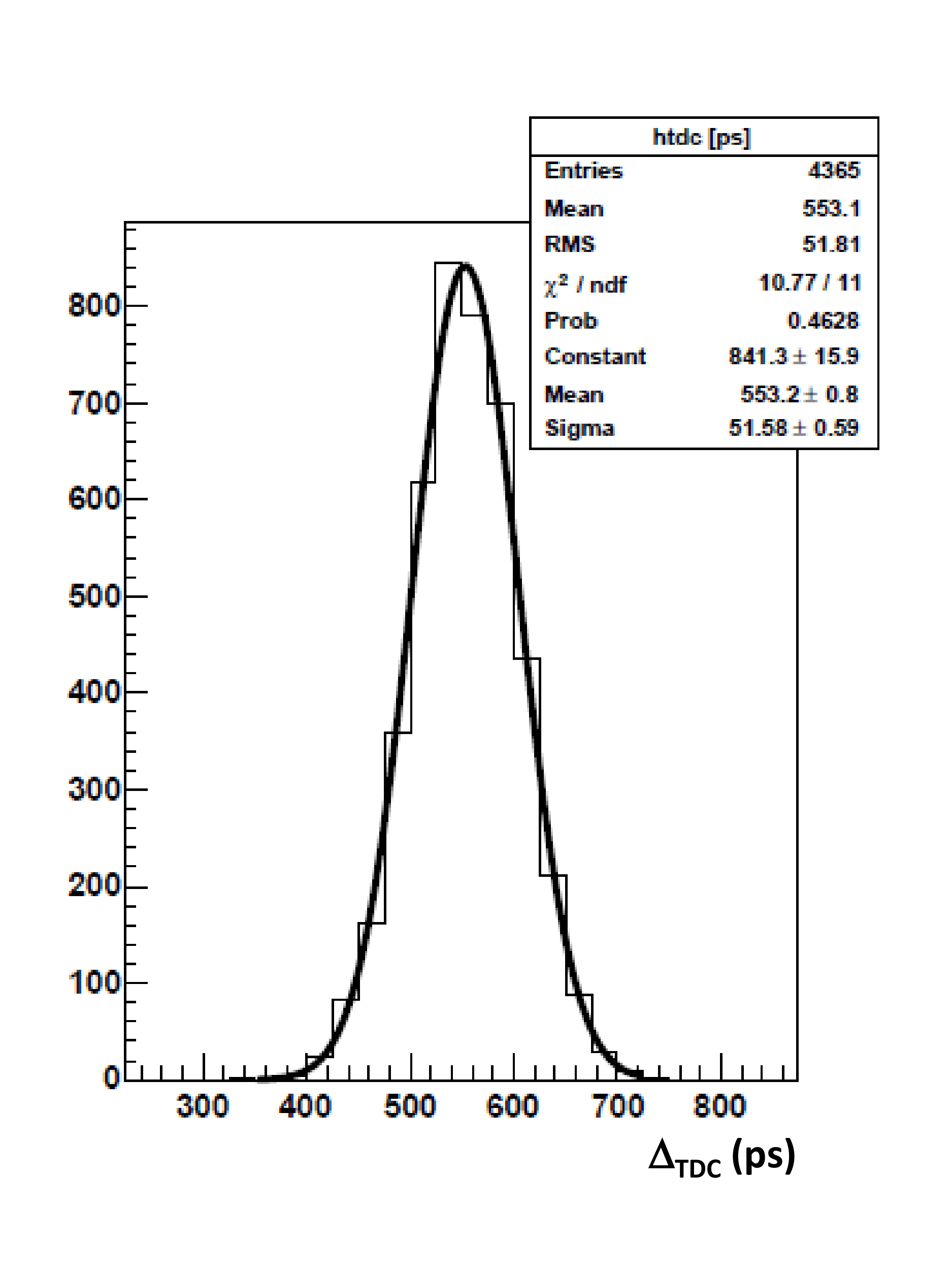}
  \caption{Distribution of $\Delta(TDC)$ in ps for the direct injection of the laser signal in fiber 
   (left panel) 
   and after the calibration system (right panel).}
  \label{fig:calib}
 \end{figure}

The intrinsic detector resolution $\sigma_t$ may be evaluated from a gaussian
fit to the $\Delta(TDC)=TDC_L-TDC_R$ distribution, as 
$\sigma_t=\sigma_{\Delta(TDC)}/2$, see \cite{villaolmo} for more details. 
From measurements with the full calibration system inserted $\sigma(TDC_{L}-
TDC_{R})$ was estimated around 52 ps, as seen in figure \ref{fig:calib}. 
Injecting directly the laser pulse into 
the scintillator bar under test 
we obtained instead 
$\sigma(TDC_{L}-TDC_{R}) \sim 37$ ps.

Assuming that the difference is due only to the presence of the calibration 
system from the initial 1x2 splitter  to the final 1m long injection fiber, 
a calibration resolution
around 18 ps may be estimated~\footnote{ The calibration resolution 
$\sigma_{cal}$ has simply been estimated from the quadratic difference 
of $\sigma_{\Delta TDC/2}$, corresponding to the measured scintillator
counter time resolution $\sigma_t$, with and without the calibration system
inserted}.  
This number has to be compared with the intrinsic
resolution of the TOF system to be calibrated (see formula \ref{eq3}).
 and demonstrates
that such a system may be useful for TOF detectors with an intrinsic resolution
down to 50-100 ps.

\section{Conclusion}

Optical components to assemble a calibration system, based on a laser diode
as a light source, have been extensively tested.
Optical switches and  fused splitters introduce minimal deterioration on the
timing properties of the delivered laser pulse. The same is true for optical
fiber patches, if their length is less than 10-20 meters. A prototype system
has been assembled in laboratory at INFN Sezione Milano Bicocca, showing
that calibration resolutions around 20-30 ps are within reach. Therefore,
we  conclude that such a calibration system may be used for fast
TOF system based on scintillators, with up to 100-200 channels and timing
resolutions in the range 50-100 ps, such as the one developed for the
MICE experiment at RAL \cite{Gregoire:2003}

The obtained results on optical properties of fiber patches, fused fiber 
splitters, optical switches may be of interest also for the calibration/
monitoring of new timing
systems based on PMTs, such as the ones foreseen for future large Liquid 
Argon TPCs \cite{wa104}.
\section*{Acknowledgements}
We would like to acknowledge the skilfull work of Mr. R. Mazza, F. Chignoli
of INFN Milano Bicocca and M. Prata of INFN Pavia for 
help in the realization of the test setup and thank 
Dr. L. Mariani
of dB Electronics, Dr. M. Bombonati of Hamamatsu Italia
 and Ing. G. Manusardi of Fiberlan srl for helpful discussions.

\end{document}